\documentstyle[aasms4,12pt]{article}

\begin{document}

\title{The X--ray Spectrum of the Plerionic System \\
PSR B1509$-$58/MSH 15$-$52}

\author{D. Marsden, P. R. Blanco, D. E. Gruber, W. A. Heindl, 
\\ M. R. Pelling, L. E. Peterson, R. E. Rothschild}
\affil{Center for Astrophysics and Space Sciences, University of 
California at San Diego \\ La Jolla, CA 92093} 

\author{A. H. Rots\altaffilmark{1}, K. Jahoda, D. J. 
Macomb\altaffilmark{1}}
\affil{Laboratory for High Energy Astrophysics, Code 660, NASA, 
Goddard Space Flight Center \\ Greenbelt, MD 20771}
\altaffiltext{1}{Universities Space Research Association} 

\begin{abstract}

We present the results of observations of the PSR B1509$-$58/MSH 
15$-$52 system in X--rays ($2-250$ keV) by the {\it Rossi 
X--ray Timing Explorer}. The spectra of the peak of the pulsed 
component (radio phase $0.17-0.53$) is fit by a power law of 
photon index $1.36\pm0.01$, with no evidence of a high energy 
spectral break seen up to $\sim200$ keV. For the off-pulse spectral 
component, the spectrum from $2-250$ keV is fit by a power law of 
photon index $2.215\pm0.005$. An iron emission line at $6.7$ keV with 
an equivalent width of $129$ eV improves the fit, but only at a 
marginal significance. Thermal bremsstrahlung and Raymond-Smith 
models produce much worse fits to the unpulsed data. The lack of a 
high energy spectral break in the pulsed emission implies an 
efficiency of $\geq 3\%$ in the conversion of pulsar spindown 
energy to pulsed X--rays in the system.   

\end{abstract}


\section{Introduction}

PSR 1509$-$58 was discovered in X--rays by the {\it Einstein} 
satellite ($0.2-4$ keV, \cite{seward82}), and subsequent radio 
observations (\cite{manchester82}) confirmed both the $150$ ms 
period and the highest spin-down rate of any known pulsar of 
${\dot P}\sim 1.5\times10^{-12}$, which implies a rotational 
energy loss rate of $\sim2\times10^{37}$ ergs s$^{-1}$, assuming 
a neutron star moment of inertia of $10^{45}$ g cm$^{2}$. More 
recent $0.1-2.4$ keV X--ray observations by the ROSAT satellite 
(\cite{greiveldinger95}, \cite{brazier97}) revealed a 
complex morphology for the pulsar/supernova remnant system, 
possibly involving interactions via collimated outflows from 
the pulsar.

In hard X--rays, the PSR 1509$-$58/MSH 15$-$52 system was 
observed by {\it Ginga} ($2-60$ keV, \cite{kawai92}), and the 
pulsed and non-pulsed emission was modeled with power laws of 
photon indices $1.33\pm0.06$ and $2.15\pm0.02$, respectively. 
A ``weak" iron line was also reported in the phase-averaged 
{\it Ginga} spectrum, at an energy of $6.7$ keV. Observations 
at higher energies by balloon-borne instruments ($94-240$ keV, 
\cite{gunji94}), and by the BATSE instrument aboard the {\it 
Compton Gamma-Ray Observatory} ($30-800$ keV, \cite{wilson93}) 
have suggested a possible steepening of the pulsed spectrum, 
which is required for consistency with upper limits obtained at 
$\sim$ MeV (\cite{fierro95}) and $\sim$ TeV (\cite{nel93}) 
energies. 

The non-pulsed component of the high energy X--rays, presumably 
dominated by the nebular emission, has not been studied above 
$\sim60$ keV. Observations in the energy range $0.1-10.0$ 
keV by {\it ASCA} (\cite{tamura96}) have revealed a combination 
of thermal and nonthermal flux from the system, consistent with 
the picture of a compact pulsar nebula (``plerion'') surrounded 
by thermal emission from gas associated with the supernova 
remnant. The imaging spectrometers aboard {\it ASCA} showed that 
the X--ray emission at energies $\geq 2$ keV in the region of the 
supernova remnant is dominated by the flux from the pulsar and 
compact nebula.

Recently Rots et al. (1997) presented a X--ray timing analysis of 
PSR B1509$-$58 which confirmed the pulsed spectral shape and 
radio/X--ray phase lag seen by {\it Ginga}. In addition, the 
pulsed spectral index was found to be consistent with a constant 
value throughout the pulse, at odds with the behavior seen in 
hard X--rays from the Crab, another young, isolated pulsar 
(\cite{ulmer94}). 

\section{Data And Analysis}

The plerionic system PSR 1509$-$58/MSH 15$-$52  was observed by 
the High Energy X--ray Timing Experiment (HEXTE) and the 
Proportional Counter Array (PCA) instruments aboard the {\it 
Rossi X--ray Timing Explorer} ({\it RXTE}) satellite on a number 
of occasions in 1996 during both in-orbit checkout and the first 
observing period. The latter data consist of proprietary 
observations used in Rots et al. (1997). The HEXTE instrument 
consists of two clusters of collimated NaI/CsI phoswich detectors 
with a total net area of $\sim1600$ cm$^2$ and an effective 
energy range of $\sim15-250$ keV (\cite{gruber96}). HEXTE 
background estimation utilizes off-source data obtained through 
re-orientation of the detectors' viewing directions. The 
PCA instrument consists of $5$ collimated Xenon proportional 
counter detectors with a total net area of $7000$ cm$^{2}$ and 
an effective energy range of $2-60$ keV (\cite{jah96}). For the 
PCA instrument, the instrumental background estimate is 
determined from modeling of both the internal background of 
the detectors and the background due to cosmic X-ray flux and 
charged particle events.

The arrival times of the photons were corrected to the solar 
system barycenter using the JPL DE200 ephemeris and the 
source coordinates R.A. (J2000)$=15^{hr}13^{m}56.627^{s}$ and 
Dec(J2000)$=-59^{\circ}8\arcmin 9.54\arcsec$. The absolute pulse 
phase of each X--ray photon was then determined from the 
appropriate radio timing ephemeris from the ongoing monitoring 
campaign using the Parkes Telescope (\cite{kaspi94}). Description 
of the radio observations/timing ephemerides used in the analysis 
are given in Rots et al. (1997). The phase $\phi$ at time $t$ was 
obtained from the standard formula $\phi(t)={\phi}(t_{0}) + {\nu}(t-t_{0})+{1\over{2}}\dot{\nu}
{(t-t_{0})}^2+{1\over{6}}\ddot{\nu}{(t-t_{0})}^3$, where $t_{0}$ 
is the barycentric epoch corresponding to the radio timing 
ephemeris, which is defined as $t_{0geo}$ rounded to two decimal 
places in MJD. Upon determining the absolute phase of each 
photon, a folded lightcurve was obtained for each energy channel, 
relative to the appropriate radio ephemeris. Binning in pulse 
phase then produced pulsar spectra as a function of the pulse 
period. Discussion of the absolute timing accuracy of the 
instruments aboard {\it RXTE} is given elsewhere (\cite{rots97}), 
but the total uncertainty in the absolute timing is on the order 
of $\sim10$ $\mu$s, which is an insignificant ($<0.1\%$) fraction 
of the PSR B1509--58 pulse phase. 

\section{Spectral Results}

The pulse profile of PSR B1509--58 consists of a single asymmetrical 
peak below $\sim 50$ keV, which possibly develops additional components 
at higher energies (\cite{rots97}). For the spectroscopic analysis, the 
pulsed flux was taken from photons with absolute phase $0.17-0.53$, 
which encompasses the entire pulse peak, and the off-pulse flux used 
photons with phases $0.77-1.07$, corresponding to regions of the 
lightcurve away from the pulse peak (\cite{rots97}). For the PCA and 
HEXTE data, the off-pulse spectrum was obtained by subtracting the 
background model flux (PCA) or the off-source flux (HEXTE) from the 
off-pulse flux. The PCA background model was obtained using the program 
PCABACKEST, which models the time-varying  detector background, in 
addition to the constant sky background, as a function of spacecraft 
position. Because the HEXTE was in the non-rocking mode for most of 
the observations, only $\sim10\%$ of the $15-250$ keV data was available 
for the off-pulse spectral analysis. To obtain the pulsed spectrum, the 
off-pulse flux was subtracted from the pulsed flux. 
 
The PCA data and the HEXTE data were fit simultaneously to various 
spectral models using XSPEC 10.0. Because of the changing detector 
gains in the PCA between observations, the PCA data with the same 
gain were combined, and fit simultaneously to the summed, 
gain-controlled HEXTE data from each cluster. For the pulsed spectral 
fits, PCA data in the energy range $2-30$ keV were used, and for 
the off-pulse analysis PCA data from $2-20$ keV were used, because 
of uncertainties in the background model above $\sim20$ keV. HEXTE 
data in the energy range $17-250$ keV were used in both spectral fits, 
and the data from all the observation days were added (for each 
cluster). In all fits the relative normalization between the PCA and 
HEXTE was a free fitted parameter to account for uncertainties 
in the effective open area of the two instruments. The normalization 
factor of the HEXTE was found to be $60\%$ of the PCA normalization. 
For the off-pulse spectral fit, the PCA background model produced 
large residuals for much of the data in the area of the Xenon L edge 
($4-5$ keV). As a result of this, only $50\%$ of the PCA data were 
used in the spectral analysis. 

The pulsed spectrum from PSR B1509$-$58 (Figure 1) is fit equally well 
by both power law and thermal bremsstrahlung models. We adopt the 
power law model here, however, because nonthermal models are clearly 
favored in the extrapolation to higher energies (\cite{wilson93}).
The best-fit parameters for the power law fit to the pulsed spectrum 
are given in Table 1. The fitting algorithm used by XSPEC is a modified Levenberg-Marquardt routine (\cite{bev69}), and the errors for each 
parameter in Table 1 are $1\sigma$, and obtained with all the other 
parameters fixed at their best-fit values. The unpulsed $2-250$ keV 
spectrum was fit to power law, bremsstrahlung, and Raymond-Smith 
plasma models. The latter two models seriously underpredicted the high 
energy (HEXTE) X--ray flux of the unpulsed component, and the power law 
model produced the best fit to the data. The addition of a narrow gaussian 
line at $\sim6.7$ keV improved the fit, with an FTEST significance of 
$1\%$. Because of systematic uncertainties in the PCA background model 
at energies below $\sim3$ keV, the intervening column density was set to 
zero in the unpulsed spectral fits. This had a minimal effect on the 
results, due to the insensitivity of the PCA to small absorption columns 
($N_{H}<10^{22}$ cm$^{-2}$). Figure 1 shows the counts spectrum of the 
unpulsed data, fitted to the model with the gaussian line. The apparent absorption feature at $\sim 5$ keV is due to imperfect modeling of Xenon 
lines in the instrument response and background model. The best-fit 
parameters of the power law and gaussian line model and the single power 
law model for the unpulsed data are given in Table 1.

\section{Discussion}

The best-fit value of the low energy neutral hydrogen absorption 
obtained in fitting the pulsed component is $N_{H}=(1.27\pm0.23)\times
10^{22}$ cm$^{-2}$, which is higher than the value of $(0.59\pm0.06)
\times10^{22}$ cm$^{-2}$ obtained by {\it ASCA} for the thermal cloud 
only $\sim8'$ from the pulsar position (\cite{tamura96}). To 
investigate the possible inconsistency between the pulsed $N_{H}$ and 
the lower value, the best-fit contours were calculated as a function 
of $N_{H}$ and the photon index using the ``error'' routine of XSPEC, 
which calculates the contours of the fit statistic as a function of 
two parameters, while holding the other parameters fixed. The results 
indicate that a value as low as $N_{H}\sim0.6\times10^{22}$ cm$^{-2}$ 
is still allowable at the $99\%$ confidence level. The {\it Ginga} 
value (\cite{kawai92}) for the photon index and $N_{H}$ may also 
be consistent with the {\it RXTE} best-fit at the $99\%$ level. 
The {\it RXTE} value for $N_{H}$ is also roughly consistent with 
estimations based on the radio dispersion measure of the pulsar. 
Assuming $10$ neutral hydrogen atoms per free electron along the 
line of sight (\cite{saito97}), we obtain a value of 
$N_{H}\sim0.8\times10^{22}$, given a distance of $4.2$ kpc 
(\cite{clark77}) and a dispersion measure of $253$ cm$^{-3}$ pc
(\cite{kaspi94}). We conclude that the {\it RXTE} does not 
exclude $N_{H}$ values that are consistent with the interstellar 
absorption and pulsar dispersion measure. 

The existence of a spectral break in the hard X--ray spectrum of 
the pulsed emission from PSR B1509$-$58, while not seen in the 
{\it RXTE} data, is implied by observations taken at higher energies 
by EGRET (\cite{fierro95}). Figure 3 shows the best-fit {\it RXTE} 
pulsed spectrum of PSR B1509$-$58 and its extrapolation to the 
EGRET energy range. Overplotted are the $5$ BATSE data points 
(\cite{wilson93}) from $30-800$ keV, and the best-fit spectrum 
from the $94-240$ keV balloon observations of Gunji et al. (1994). 
Also shown is the {\it RXTE} unpulsed (nebular) spectrum. Aside from a 
factor of $\sim1.5$ in BATSE/{\it RXTE} normalization, the pulsed 
spectrum roughly maintains its shape out to at least $\sim400$ keV. 
The extrapolation to the EGRET energy range ($30-100$ MeV), however, 
exceeds the $2\sigma$ EGRET upper limit (\cite{fierro95}) by a factor 
of $10-100$, necessitating a break in the pulsed spectrum 
somewhere in the energy range $\sim0.4-30$ MeV.

Because the {\it RXTE} pulsed photon index is $<2$, the integrated 
X--ray energy flux of the pulsar is actually increasing with increasing 
photon energy, putting constraints on the efficiencies of the processes 
converting the spindown energy of the neutron star into observable 
radiation. If we assume the pulsar and plerionic flux are due to 
conversion of spindown energy with efficiencies ${\epsilon}_{p}$ 
and $\epsilon_{op}$, respectively, this constraint can be expressed 
as:
\begin{equation}
L_{spin}\geq4\pi{d^{2}}\int_{E_{0}}^{E_{b}}(\eta{\epsilon_{p}}^{-1}
F_{p}+{\epsilon_{op}}^{-1}F_{op})dE
\end{equation}
where $L_{spin}$ and $d$ are the pulsar spindown luminosity 
($I\Omega{\dot \Omega}$) and distance, respectively. In Equation (1), 
$F_{p}$ and $F_{op}$ are the energy fluxes from the best fit pulsed 
and unpulsed spectral models from the {\it RXTE} data, and $\eta$ is 
the solid angle fraction subtended by the pulsar beam (the unpulsed, 
plerionic emission is assumed to be isotropic). For the pulsed 
beaming fraction we adopt the value $\eta=0.3$, which is 
estimated from calculations by Chiang \& Romani (1992). Assuming 
efficiencies that are independent of energy, and a distance of 
$4.2$ kpc to the pulsar, and setting the lower energy $E_{0}=2$ keV 
and $L_{spin}=1.8\times10^{37}$ ergs s$^{-1}$, yields values for the 
efficiencies as a function of break energy $E_{b}$ given by Figure 4. 
The calculation in Figure 4 assumes that the entire spindown energy 
budget of the pulsar is expended between the energies of $E_{0}$ and 
$E_{b}$. Extension of the {\it RXTE} pulsed and unpulsed spectra to the 
radio point at $1.5$ GHz (\cite{manchester82}) implies a total 
contribution to the spindown energy budget from photons below $E_{0}$ 
of $<3\%$ of the $2-250$ keV flux, given the radio flux density of 
$2$ mJy. 

The break energies in Figure 4 are relatively insensitive to the 
assumed spectral flattening at energies below $\sim2$ keV, because 
the photons at these energies carry $<3\%$ of the energy of the 
photons in the {\it RXTE} bandpass. Assuming a plerionic efficiency of 
${\epsilon}_{op}=10-20\%$ (see below), a pulsed efficiency of 
${\epsilon}_{p}\geq3\%$ is required to explain the lack of a spectral 
break below $\sim400$ keV implied by the combined {\it RXTE} and BATSE observations. A plerion efficiency of ${\epsilon}_{op}<10\%$ would 
require an even higher pulsed efficiency, but this could be eased if 
the pulsar has a narrow beam ($\eta<0.3$).

The pulsed efficiency obtained above is slightly higher than the 
$1-3\%$ efficiency predicted by outer gap models of high energy 
emission from young pulsars (\cite{cheng86}). Outer gap models have 
difficulties explaining the lack of GeV gamma--rays from PSR B1509$-$58 
as well, because in these models the emission sites are located in the 
outer magnetosphere, where attenuation due to magnetic effects should 
be minimal (\cite{chang96}). A polar cap model of the PSR B1509$-$58 
pulsed emission (\cite{harding97}) explains the attenuation of the 
gamma--ray photons in terms of magnetic photon splitting and pair 
production, but places the spectral break at energies $\geq2$ MeV, 
which would require a relatively large pulsed efficiency of at least 
$7\%$ (Figure 4). This model also requires a polar cap for PSR 
B1509$-$58 much larger than produced by a simple dipole field geometry (\cite{chang96}), indicating that perhaps a more complicated magnetic 
field geometry is required to explain the pulsed spectrum. 

The spectral fits indicate that a power law form is clearly favored 
for the unpulsed spectrum, and dominates the total X--ray emission 
from the PSR B1509$-$58/MSH 15$-$52 system at energies $\leq10$ keV 
(Figure 3). The {\it ASCA} images (\cite{tamura96}) indicate that the 
$\sim2-10$ keV emission from the system is concentrated around 
the pulsar, and not the surrounding supernova remnant, implying 
that the unpulsed $2-250$ keV emission seen by {\it RXTE} is due to 
the pulsar plerion, and not supernova emission from MSH 15$-$52. The 
high energy nonthermal emission from plerions is believed to be 
synchrotron radiation from a luminous ``bubble'' confined by the 
pressure of the surrounding supernova remnant (\cite{rees74}). The 
internal pressure of the bubble is provided by a wind of particles 
and magnetic fields ejected by the pulsar. The $2-250$ keV luminosity 
of the PSR B1509$-$58 plerion is $L_{x}=(4.7\pm0.9)\times10^{35}$ 
ergs s$^{-1}$, assuming a distance of $4.2$ kpc to the source. 

Detailed models of the Crab Nebula (\cite{kennel84}, \cite{kennel284}) 
indicate that an efficiency of $10-20\%$ can be achieved using 
reasonable values for the pulsar wind speed, plerion size, and ratio 
of electromagnetic to particle energy in the wind. For the PSR 
B1509$-$58/MSH 15$-$52 system, Figure 4 indicates that similar plerionic 
efficiencies are allowed if the pulsed efficiency is not too high 
($\epsilon_{p}\leq3\%$). A higher pulsed efficiency would require a 
lower plerionic efficiency, as mentioned above. The $2-250$ keV 
spectral index obtained for the PSR B1509$-$58 plerion is slightly 
steeper than the $17-180$ keV index obtained for the Crab Nebula 
(\cite{jung89}), but the latter index show evidence for gradual 
steepening throughout the X--ray to gamma--ray energy range, so the 
two indices may be consistent. 

The marginal detection of a line at $\sim6.7$ keV suggests emission 
from H--like iron. The source of the iron line is most likely either 
the supernova remnant/plerion or the galactic ridge. The X--ray emission 
from the latter was mapped out by the Japanese satellite {\it Tenma}, 
which found a mean iron line centroid of $\sim6.7$ keV that varied 
in both energy and flux as a function of position (\cite{koyama89}).
The {\it RXTE} line centroid is consistent with the galactic ridge 
emission seen by {\it Tenma}, but the flux is $\sim3$ times greater 
than expected from the galactic ridge, suggesting a contribution from 
iron in MSH 15$-$52.  

\section{Conclusions}

Observations of the PSR B1509$-$58/MSH 15$-$52 system by {\it RXTE} 
suggest that the hard X--ray spectrum from the system is dominated by 
nonthermal emission. The spectral shape of the pulsed and unpulsed 
radiation are characterized by power laws of photon index $1.358\pm0.014$ 
and $2.215\pm.005$, respectively. Considerations of the {\it RXTE} data 
in the context of observations of PSR B1509$-$58/MSH 15$-$52 at other 
wavelengths allow us to constrain the efficiencies of the conversions 
of pulsar spindown energy to observed radiation in the system. The lack 
of a significant spectral break in the X--ray emission out to hundreds 
of keV allows the efficiencies for the conversion of particle 
energy to X--rays in the pulsar magnetosphere and compact nebula to be constrained. The minimum pulsed efficiency is found to be $3\%$, assuming 
an efficiency for the unpulsed radiation similar to that of the Crab 
Nebula. 

\acknowledgments

We acknowledge fruitful discussions with Richard Lingenfelter and 
a critical reading of the manuscript by Vicky Kaspi. The up-to-date 
radio timing ephemeris was provided by M. Bailes, V. Kaspi and R. 
Manchester using data obtained at the Parkes radio telescope in 
Australia. This work was funded by NASA grant NAS5-30720.

\clearpage

\begin{figure}
\plotfiddle{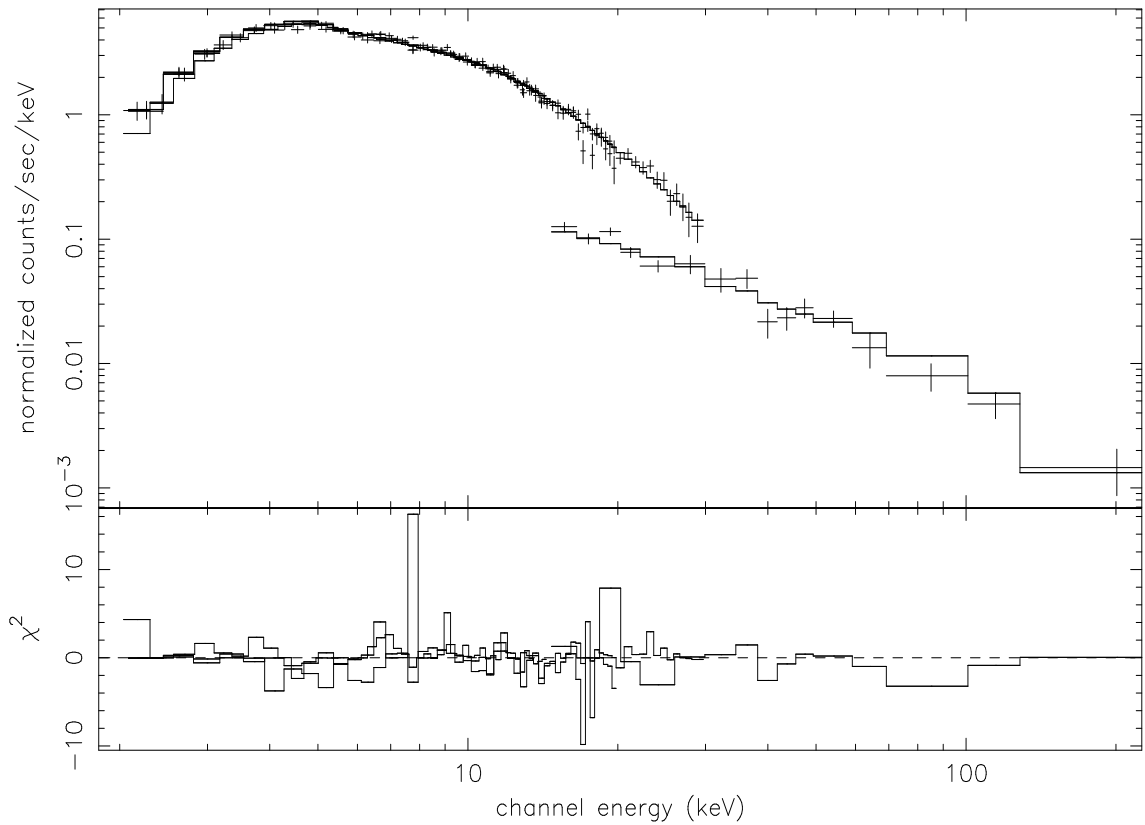}{6in}{-90.}
{65.}{70.}{-252pt}{396pt}
\caption{~The spectral fit of the pulsed data to the model 
given in Table 1, showing the counts data with the 
fitted model overplotted as a solid line (above), 
and the residuals (below). The PCA data ($2-30$ keV) have 
been multiplied by a factor of $0.6$ to account for 
uncertainties in the effective area calibrations of PCA 
and HEXTE. The HEXTE data have been co-added for clarity.}
\end{figure} 

\begin{figure}
\plotfiddle{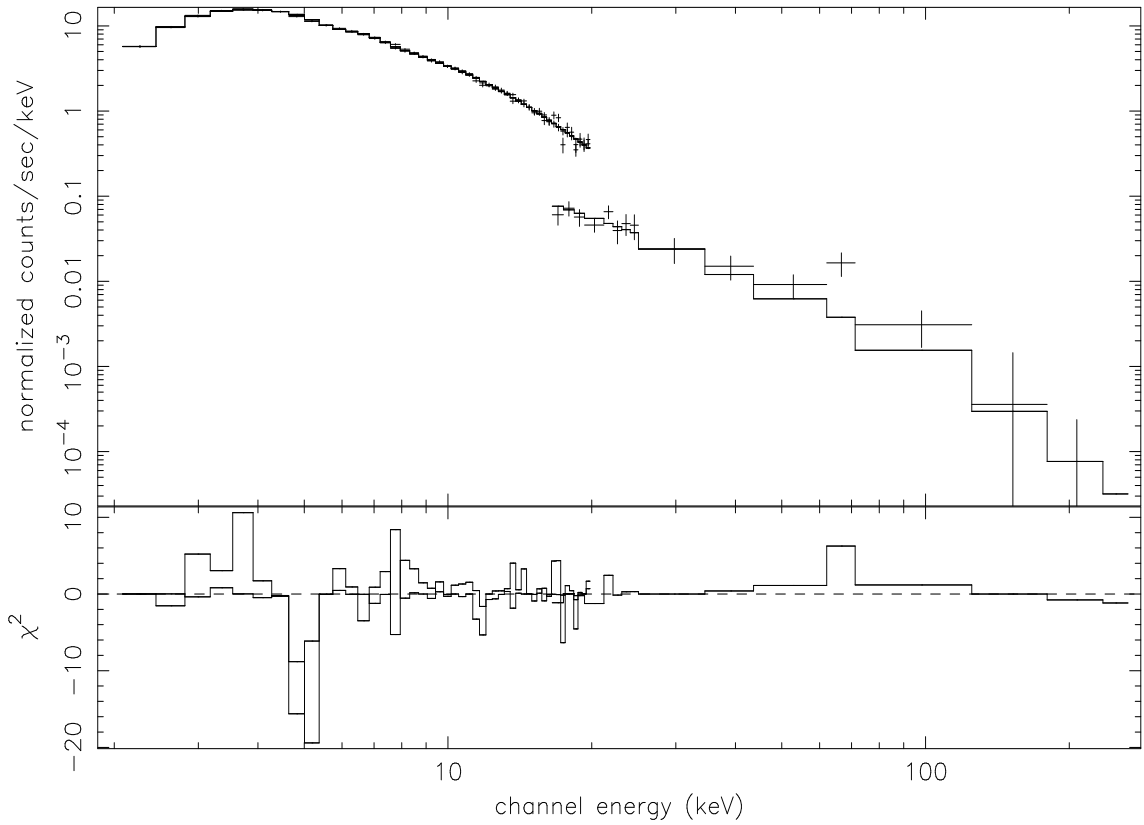}{6in}{-90.}
{65.}{70.}{-252pt}{396pt}
\caption{~Same as Figure 1, but with the off-pulse data. 
Addition of a narrow gaussian line at $6.7$ keV to the power 
law model improved the fit significantly. The feature at 
$\sim5$ keV is a result of oversubtraction of Xenon lines 
in the background model.} 
\end{figure}

\begin{figure}
\plotone{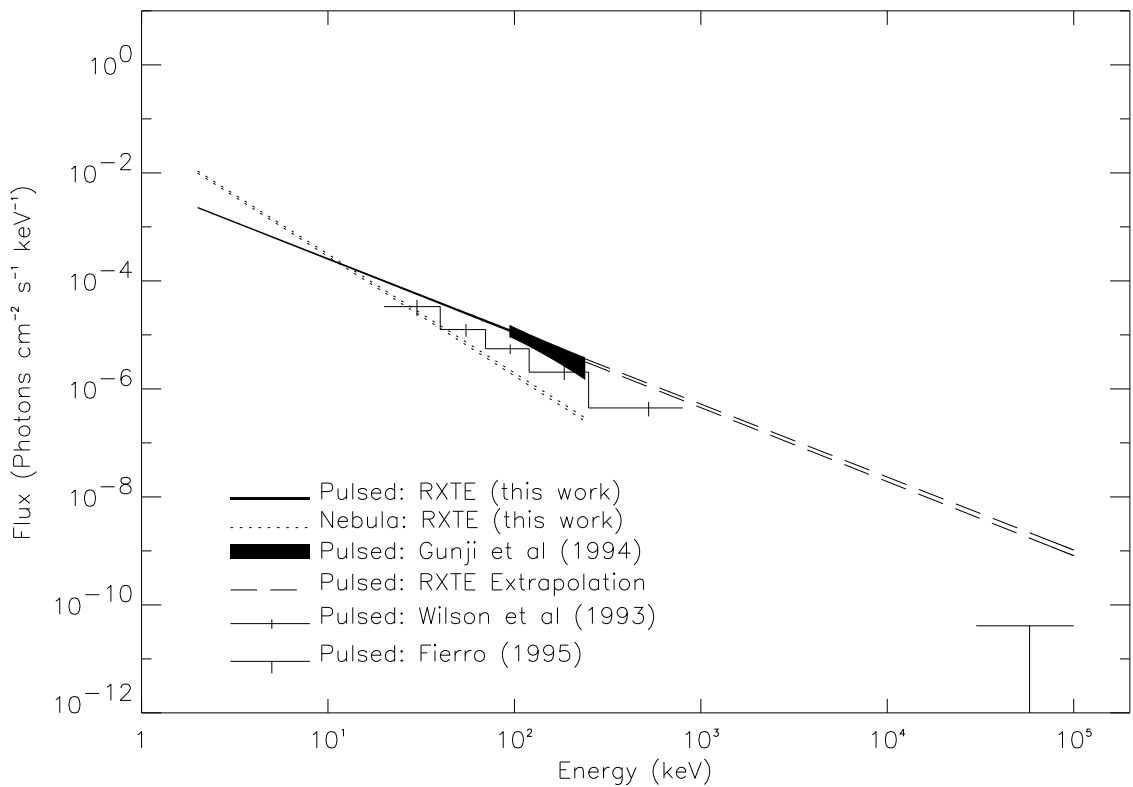}
\caption{~The broadband high energy photon spectrum of 
PSR B1509$-$58/MSH 15$-$52. The dashed lines encompass the $1\sigma$ 
error region of the extrapolated {\it RXTE} pulsed spectrum, and 
the dotted lines show the best--fit photon spectrum for the unpulsed 
data over the {\it RXTE} energy range. The extrapolated pulsed 
spectrum from PSR B1509$-$58 overpredicts the EGRET gamma--ray flux 
by $1-2$ orders of magnitude.}
\end{figure}

\begin{figure}
\plotone{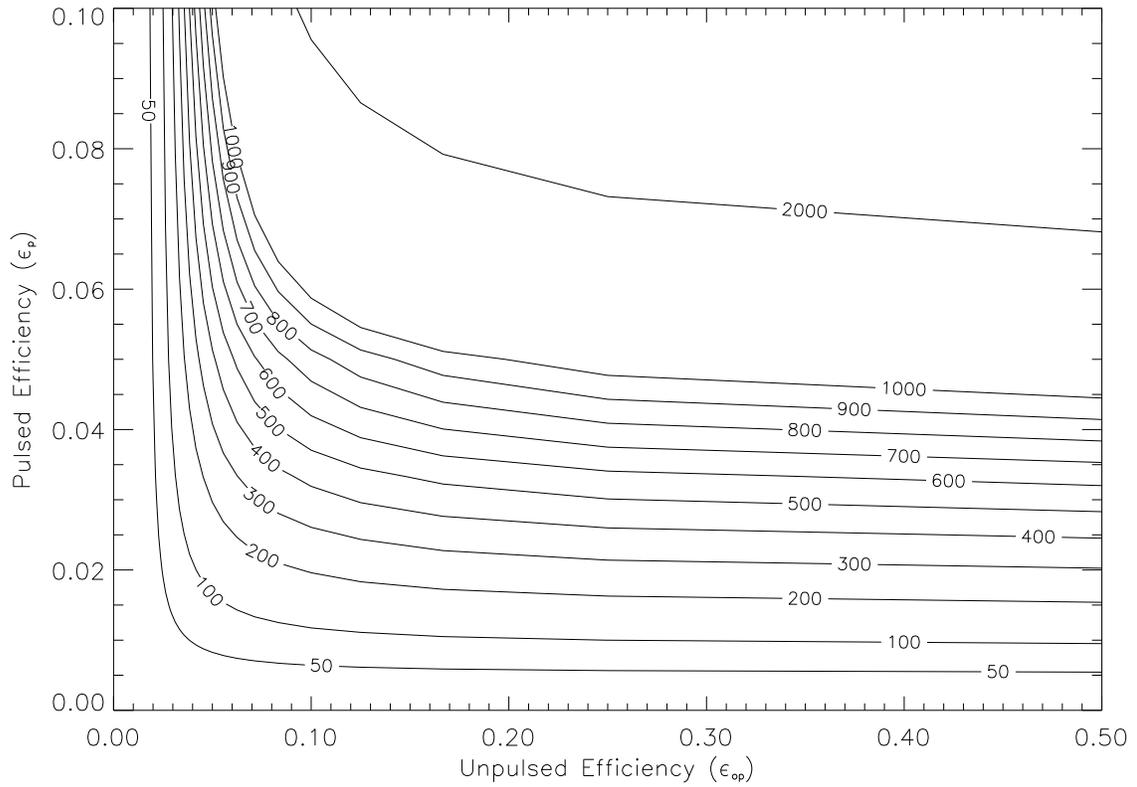}
\caption{~The upper break energy of the pulsed spectrum, shown as 
a function of the pulsed and unpulsed efficiencies. The 
calculation assumes a lower break energy of $2$ keV and a 
beaming factor of $0.3$ (see text). The break energies are in keV.}
\end{figure}

\clearpage

\begin{deluxetable}{lccc}
\footnotesize
\tablewidth{0pt}
\tablecaption{Spectral Fit Results \label{tbl-2}}
\tablehead{\colhead{Model Parameter} & \colhead{Pulsed Emission} & 
\colhead{Off-Pulse Emission} & \colhead{Off-Pulse Emission}} 
\startdata
Normalization\tablenotemark{a} & $6.07\pm0.21$ & $52.88\pm0.43$ & 
$53.55\pm0.44$ \nl
Photon Index & $1.358\pm0.014$ & $2.200\pm0.005$ & $2.215\pm0.005$ \nl
N$_{H}$\tablenotemark{b} & $1.27\pm0.23$ & $0$ (fixed) & $0$ (fixed) \nl
E$_{l}$\tablenotemark{c} & $-$ & $-$ & $6.71\pm0.05$ \nl
N$_{l}$\tablenotemark{d} & $-$ & $-$ & $1.77\pm0.27$ \nl
${\chi_{\nu}}^{2}$ & $0.99$ & $1.34$ & $1.14$ \nl
$\nu$ & $862$ & $763$ & $761$\nl
\enddata
\tablenotetext{a}{$10^{-3}$ Photons cm$^{-2}$ s$^{-1}$ kev$^{-1}$ at 
$1$ keV}
\tablenotetext{b}{Neutral hydrogen absorption ($10^{22}$ H Atoms cm$^{-2}$)}
\tablenotetext{c}{Line centroid energy (keV)}
\tablenotetext{d}{Total flux in line ($10^{-4}$ Photons cm$^{-2}$ s$^{-1}$)}
\end{deluxetable}

\end{document}